\documentclass[conference]{IEEEtran}
\IEEEoverridecommandlockouts

\usepackage{cite}
\usepackage{amsmath,amssymb,amsfonts}
\usepackage{algorithmic}
\usepackage{graphicx}
\usepackage{textcomp}
\usepackage{xcolor}
\usepackage{booktabs}
\usepackage{float}
\usepackage{microtype}
\usepackage{flushend}

\def\BibTeX{{\rm B\kern-.05em{\sc i\kern-.025em b}\kern-.08em
    T\kern-.1667em\lower.7ex\hbox{E}\kern-.125emX}}
\begin{document}

\title{Input-Correlated Supervision Noise Limits the Benefits of OTA Training for Learned Receivers}

\author{\IEEEauthorblockN{Riku Luostari}
\IEEEauthorblockA{\textit{Nokia} \\
Espoo, Finland \\
riku.luostari@nokia.com}
\and
\IEEEauthorblockN{Dani Korpi}
\IEEEauthorblockA{\textit{Nokia Bell Labs} \\
Espoo, Finland \\
dani.korpi@nokia-bell-labs.com}
\and
\IEEEauthorblockN{Olav Tirkkonen}
\IEEEauthorblockA{\textit{Aalto University} \\
Espoo, Finland \\
olav.tirkkonen@aalto.fi}
\and
\IEEEauthorblockN{Harri Holma}
\IEEEauthorblockA{\textit{Nokia} \\
Espoo, Finland \\
harri.holma@nokia.com}
\thanks{Accepted for publication in the IEEE GLOBECOM 2026 Workshops.
\copyright~2026 IEEE. Personal use of this material is permitted. Permission
from IEEE must be obtained for all other uses, in any current or future media,
including reprinting/republishing this material for advertising or promotional
purposes, creating new collective works, for resale or redistribution to servers
or lists, or reuse of any copyrighted component of this work in other works.}
}

\maketitle
\thispagestyle{plain}
\pagestyle{plain}

\begin{abstract}
        While learned wireless receivers are typically studied using synthetic data, the impact of over-the-air (OTA) measurements for training remains unclear. We conducted a 5.88 GHz measurement campaign with a 5G/6G-like orthogonal frequency-division multiplexing (OFDM) system across diverse environments and mobility conditions, and trained a neural channel estimator and a capacity-matched end-to-end neural receiver using mixtures of measured and synthetic data. Increasing the OTA fraction revealed a fundamental asymmetry: measured data consistently improved the end-to-end receiver, whereas the channel estimator peaked at an intermediate fraction and degraded with fully measured training. We identified supervision quality as a key contributor: OTA channel labels are derived from noisy received signals and therefore contain supervision errors correlated with the receiver input, whereas decoded bits validated by a cyclic redundancy check (CRC) provide effectively error-free supervision. Denoising the channel labels mitigated the degradation, and a controlled simulation showed that label noise of equal power degrades the estimator far more when it is correlated with the input. For OTA training, end-to-end receivers thus benefit from fully measured data, whereas channel estimators benefit from moderate OTA fractions but require improved label quality.

\end{abstract}

\begin{IEEEkeywords}
Channel Estimation, Neural Receiver, Deep Learning, OFDM, Over-the-Air Measurements.
\end{IEEEkeywords}

\section{Introduction}

Deep learning has been applied across the wireless receive chain, from modular channel estimation to end-to-end receivers such as DeepRx that map received signals directly to bit log-likelihood ratios (LLRs) \cite{hiray_neural_2016,mei_low_2021,honkala_deeprx_2021,pihlajasalo_hybriddeeprx_2021}. These approaches have been developed primarily for orthogonal frequency-division multiplexing (OFDM) systems, including those considered for 6G \cite{3gpp_study_2026,3gpp_feature_2025}. In conventional receivers, channel estimation relies on pilots and model-based methods such as least-squares (LS), minimum mean square error (MMSE) interpolation, and decision-directed refinement \cite{coleri_channel_2002,morelli_comparison_2001,mignone_cd3-ofdm_1996,sanzi_comparative_2003,komninakis_multi-input_2002}. Learned receivers instead infer channel characteristics, hardware effects, and detection rules directly from data. However, their training and evaluation have almost exclusively relied on synthetic channels, leaving the practical value of over-the-air (OTA) measurements insufficiently understood \cite{luostari_adapting_2025}. Independently, Baytekin et al.\ have studied site-specific fine-tuning of neural receivers with OTA measurements from a 5G system \cite{baytekin_site-specific_2026}.

In this work, we show that the value of OTA data depends strongly on the learning objective. The same measurements that consistently improve an end-to-end neural receiver can degrade a learned channel estimator beyond a certain OTA fraction. We identify input-correlated supervision noise as a key mechanism: OTA channel labels are constructed from noisy received observations and therefore contain a noise component correlated with the network input. Unlike independent label noise, the resulting bias cannot be removed by collecting more measurements.

\begin{figure}[t]
    \centerline{\includegraphics[width=\columnwidth]{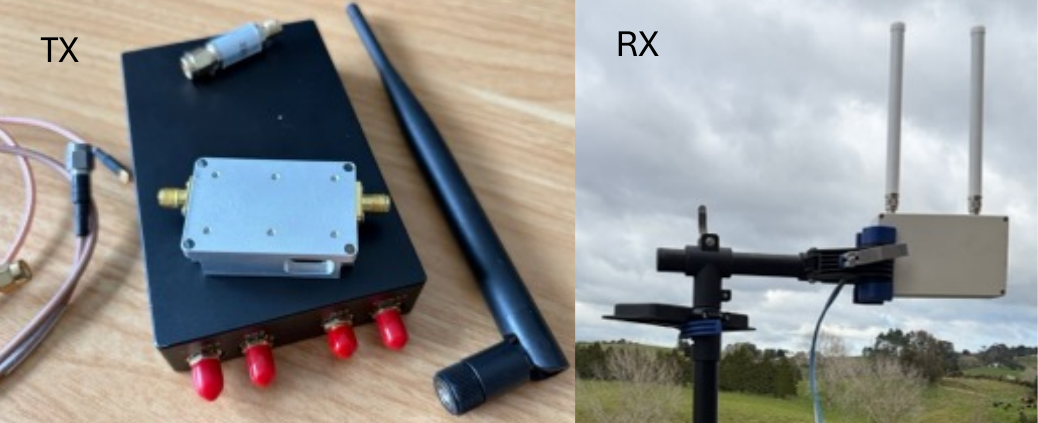}}
    \caption{Over-the-air measurement hardware, based on the ADALM-PLUTO software-defined radio (SDR) \cite{collins_software-defined_2018}: the transmitter with external power amplifier, filter, and antenna (left) and the base-station receiver with antenna (right).}
    \label{fig:measkit}
    \end{figure}

To investigate this effect, we conducted a 5.88~GHz measurement campaign spanning diverse environments and mobility conditions, using a 5G/6G-like OFDM waveform. We trained two approximately capacity-matched receivers with controlled mixtures of simulated and measured data: CM DeepRx (ConvMixer DeepRx), an end-to-end LLR receiver, and ACM-E (Axial ConvMixer Estimator), a neural channel estimator in a conventional receiver chain. Matching their parameter counts reduced model capacity as a confound but did not remove differences in learning objective or receiver structure. We further tested the proposed mechanism by denoising OTA channel labels while holding the training inputs and architecture fixed, and in a controlled simulation.

We restricted the study to single-input single-output (SISO) operation because the dense symbol-normalized channel labels used here do not directly extend to multiple-input multiple-output (MIMO). Nevertheless, the underlying issue applies more generally whenever supervision targets are derived from noisy received observations.

\textbf{Contributions.} \emph{(i)} We show that increasing OTA training data produces fundamentally different behavior for end-to-end receivers and neural channel estimators. \emph{(ii)} We identify input-correlated supervision noise in OTA channel labels and show theoretically why additional measurements cannot eliminate its bias; a label-denoising experiment and a controlled simulation support this mechanism. \emph{(iii)} We derive practical guidance for OTA training: end-to-end receivers benefit from increasing measured data, whereas intermediate estimators require improved supervision quality rather than simply more measurements.

\section{System Model}

We consider a SISO OFDM system with $N_\mathrm{FFT}$ subcarriers and $N_s$ symbols per slot. The received signal is
\begin{equation}
y_{ij} = H_{ij} x_{ij} + n_{ij},
\end{equation}
where $x_{ij}$ is the transmitted quadrature amplitude modulation (QAM) symbol, $H_{ij}$ the channel gain, and $n_{ij} \sim \mathcal{CN}(0,\sigma_n^2)$.
The indices $i$ and $j$ denote the OFDM symbol and active subcarrier, respectively.

After removing guard bands and the DC subcarrier, $N_\mathrm{eff}$ subcarriers carry data or pilots. Pilot symbols are located on indices $\mathcal{P} \subset \{1,\dots,N_s\} \times \{1,\dots,N_\mathrm{eff}\}$.

At pilot positions, the LS estimate is
\begin{equation} \label{eq:ls}
\hat{H}^{\mathrm{LS}}_{ij} = \frac{y_{ij}}{x_{ij}} = H_{ij} + \tilde{n}_{ij}, \quad (i,j)\in\mathcal{P},
\end{equation}
where $\tilde{n}_{ij} \triangleq n_{ij}/x_{ij}$ is the effective noise term with variance $\sigma_n^2/|x_{ij}|^2$. Channel values at non-pilot positions are obtained via bilinear interpolation of $\hat{H}^{\mathrm{LS}}_{ij}$ over $\mathcal{P}$ \cite{coleri_channel_2002,morelli_comparison_2001}.

\section{Methods}

\subsection{Iterative DDCE: Decision-Directed Channel Estimation} \label{sec:ddce} As a non-learned reference, we used a three-iteration decision-directed channel estimation (DDCE) scheme initialized by the LS estimate \cite{mignone_cd3-ofdm_1996,sanzi_comparative_2003}. Each iteration performs zero-forcing (ZF) equalization, hard detection, symbol-normalized channel reconstruction, and separable uniform time-frequency smoothing with kernels ($k_f = 5$, $k_t = 3$) chosen for this work.

\begin{table}[H]
\caption{Learned Receiver Models}
\label{tab:models}
\centering
\begin{tabular*}{\columnwidth}{@{\extracolsep{\fill}}llll@{}}
\toprule
\textbf{Model} & \textbf{Input} & \textbf{Output} & \textbf{Loss (target)} \\
\midrule
ACM-E     & $\mathbf{Y}$, pilots & Estimate $\hat{\mathbf{H}}$ & MSE ($\mathbf{H}^\mathrm{gt}$) \\
CM DeepRx & $\mathbf{Y}$, pilots & Bit LLRs                    & BCE (bits) \\
\bottomrule
\end{tabular*}
\end{table}

\subsection{ACM-E: Axial ConvMixer Estimator} 

We developed ACM-E, a neural channel-estimation architecture that adapts axial factorization and ConvMixer blocks from image processing \cite{ho_axial_2019,trockman_patches_2022} to OFDM channel estimation. Independently of this work, a similar separable factorization is used in the neural receiver of Yellapragada et al.~\cite{yellapragada_computationally_2026}. ACM-E produces a channel estimate $\hat{\mathbf{H}}$ for a conventional ZF equalizer and demapper (Table~\ref{tab:models}). Unlike LS-based methods, its full-signal input $\mathbf{Y}$ enables learning of constellation structure and hardware effects. Axial factorization replaces 2D convolutions with cascaded 1D depthwise convolutions along time and frequency, providing a wide receptive field with low complexity and generalizing DDCE's fixed smoothing kernels. The encoder uses six Axial ConvMixer blocks ($d=128$, $K_t=3$, $K_f=5$) over two resolution stages, followed by a bottleneck ($d_p=16$) and decoder with skip connections.

Training minimizes the mean square error (MSE) reconstruction loss
\begin{equation} \label{eq:recon_loss}
\mathcal{L}_\mathrm{ACM\text{-}E}
=\frac{1}{N_sN_\mathrm{eff}}\sum_{i,j}
|\hat{H}_{ij}-H^\mathrm{gt}_{ij}|^2 .
\end{equation}
The target $\mathbf{H}^\mathrm{gt}$ is the genie channel for simulation and $H^\mathrm{gt}_{ij}=y_{ij}/x_{ij}$ for OTA data. The latter is unbiased for $H_{ij}$ with variance inversely proportional to the signal-to-interference-plus-noise ratio (SINR), but its error is correlated with $\mathbf{Y}$, biasing the learned predictor rather than the label itself (Sec.~\ref{sec:label_noise}).

ACM-E was trained for 100\,000 iterations with batch size 32.

\subsection{CM DeepRx Neural Receiver} \label{sec:deeprx} For this study, we developed CM DeepRx, which, in contrast to ACM-E, jointly optimizes channel estimation, equalization, and demapping end-to-end. It is a ConvMixer-based variant of the DeepRx framework \cite{honkala_deeprx_2021}, replacing the original ResNet backbone \cite{he_deep_2016} with $L=6$ ConvMixer blocks \cite{trockman_patches_2022} of hidden dimension $d=256$ and asymmetric depthwise kernels ($K_t=3$, $K_f=5$). This practical variant converged substantially faster in our experiments while preserving similar performance. Like ACM-E, it takes $\mathbf{Y}$ and pilot information as input, but directly produces per-bit LLRs (Table~\ref{tab:models}) without an explicit channel-estimation stage.

The wider hidden dimension ($d=256$ versus $d=128$ for ACM-E) compensates for ACM-E's deeper architecture, yielding comparable model capacities that differ by approximately 15\% (Table~\ref{tab:complexity}). Comparing architectures at similar parameter counts reduces model size as a confound but does not isolate the effects of their learning objectives and receiver structures. The output layer produces $\log_2 M$ LLRs per resource element, where $M$ is the modulation order. Training used the binary cross-entropy (BCE) loss for 400\,000 iterations with batch size 8,
\begin{equation} \label{eq:bce_loss}
\mathcal{L}_\mathrm{DeepRx} = -\frac{1}{N_b}\sum_k \left[ b_k \log \hat{p}_k + (1-b_k)\log(1-\hat{p}_k) \right],
\end{equation}
where $b_k$ is the transmitted bit, $\hat{p}_k$ the predicted bit probability, and $N_b$ the number of bits.
The iteration counts followed observed convergence; the differing batch sizes gave both models 3.2 million training samples.

\subsection{Measurement Campaign}
\label{sec:measurement}

Measurements were conducted at 5.88~GHz using modified PlutoSDR software-defined radios in a SISO configuration. The transmitter was handheld or vehicle-mounted, while the receiver was deployed at fixed base-station locations. An external power amplifier increased the transmit power to 30~dBm. The OTA data capture propagation and practical radio-frequency impairments, including nonlinear distortion, I/Q imbalance, phase noise, and synchronization errors.

The dataset covers urban, suburban, rural, and industrial environments under pedestrian and vehicular mobility (0--15~m/s). Approximately 20 sessions of 15 minutes each produced $\approx$44\,000 retained OFDM slots, which were split by session into 90\% training and 10\% held-out test data.

Each received slot was synchronized using known preambles and pilots and decoded at code rate 0.25. Only slots that passed the cyclic redundancy check (CRC) were retained. A separate acquisition counter showed that fewer than 1\% of time-synchronized slots failed this filter throughout the campaign. Because rejected slots were not stored, their SINR and scenario distributions cannot be reconstructed; consequently, CRC filtering may introduce a small selection bias toward easier realizations. For each retained sample, SINR was estimated from pilot resource elements, and the decoded bits were re-encoded and re-modulated to reconstruct the transmitted symbols $x_{ij}$, enabling dense channel-label generation through $H^\mathrm{gt}_{ij}=y_{ij}/x_{ij}$. Residual undetected decoding errors are assumed negligible.

\subsection{Mixed Simulation-Measurement Training}
\label{sec:mixed_training}

Both architectures were trained using mixed batches containing an OTA fraction $p_m \in \{0,20,\dots,100\}\%$, with the remaining samples generated synthetically. Simulated data were generated online using the 3GPP Urban Macro (UMa) channel model and provide effectively unlimited channel realizations with noise-free labels \cite{3GPP:38.901}. The goal was not to optimize the simulator but to isolate the effect of replacing synthetic data with measurements.

OTA samples were randomly drawn and augmented online with Gaussian noise on $\mathbf{Y}$ to reduce overfitting. This augmentation was applied to the input only, not the target. The original measurement noise remains shared between the input and target, but the added independent noise attenuates this dependence and can change its magnitude.

At $p_m=20\%$, approximately 640\,000 OTA draws correspond to 16 passes through the approximately 40\,000-slot training pool. Thus, every nonzero-$p_m$ setting effectively encounters all OTA slots; increasing $p_m$ changes their reuse and weight relative to online simulation.

\section{Analysis of OTA Supervision}
\label{sec:label_noise}

Although both architectures receive the same noisy observation $\mathbf{Y}$, their supervision differs. CM DeepRx is trained on CRC-validated bits, which provide effectively error-free targets (Sec.~\ref{sec:measurement}). ACM-E instead requires dense channel supervision: for simulation, the target is the genie channel $H_{ij}$, whereas OTA labels must be reconstructed from the received signal, \begin{equation} \label{eq:ota_label} H^\mathrm{gt}_{ij} = \frac{y_{ij}}{x_{ij}} = H_{ij} + \frac{n_{ij}}{x_{ij}}. \end{equation} This extends the pilot LS estimate \eqref{eq:ls} to the full resource grid, so OTA supervision carries the same measurement noise present in the input $\mathbf{Y}$, with variance $\sigma_n^2/|x_{ij}|^2 \propto 1/\mathrm{SINR}$ under the normalization used here. The argument does not require purely Gaussian noise: in OTA, $n_{ij}$ also absorbs unmodeled transceiver effects, and only the shared noise realization between input and target matters.

The MSE objective in \eqref{eq:recon_loss} converges to the conditional expectation of the target given the input:
\begin{equation} \label{eq:bayes} f^\star(\mathbf{Y})_{ij} = \mathbb{E}\!\left[H^\mathrm{gt}_{ij} \mid \mathbf{Y}\right] = \mathbb{E}\!\left[H_{ij} \mid \mathbf{Y}\right] + \mathbb{E}\!\left[\frac{n_{ij}}{x_{ij}} \mid \mathbf{Y}\right]. \end{equation}
The second term in \eqref{eq:bayes} is what distinguishes OTA supervision. For input-independent label noise, it vanishes, and more data only removes variance. Here, however, the label error $n_{ij}/x_{ij}$ is contained in the input $\mathbf{Y}$, so the term is generally nonzero and the network learns to reproduce part of the noise instead of removing it. In the limiting case where $x_{ij}$ is known, as at pilot positions, the label is fully determined by the input, so the MSE-optimal output equals the noisy label and none of the noise is removed. At data positions, $x_{ij}$ must first be inferred from $\mathbf{Y}$; once detection becomes reliable at high SINR, the same holds, so the effect grows with SINR even though the label-noise variance decreases. Additional measurements reduce variance but cannot remove the resulting bias relative to the clean channel. We term this mechanism \emph{input-correlated supervision noise}.

The OTA label \eqref{eq:ota_label} is also the unsmoothed ``divide'' step of DDCE (Sec.~\ref{sec:ddce}), except that it uses known transmitted symbols rather than detected decisions. DDCE improves this estimate through explicit noise suppression, whereas ACM-E is trained to reproduce the noisy observation-derived target, explaining why label denoising improves performance in Sec.~\ref{sec:label_denoising}.

\section{Results}
\label{sec:results}

\subsection{Experimental Setup}

System and training parameters are summarized in Table~\ref{tab:ofdm_params}. Models were trained with mixed simulated and OTA data, where $p_m$ controls the OTA fraction; evaluation was performed on held-out OTA measurements.

To account for training variability, each $p_m$ configuration was trained with $S = 5$ random seeds. Results are reported as the mean performance, with standard deviations included for the primary SINR-gain metrics. We distinguish between training \emph{SNR} (signal-to-noise ratio), the simulated noise level, and evaluation \emph{SINR}, which includes thermal noise and residual interference.

The normalized mean square error (NMSE) of the channel estimate cannot be used to compare the two receivers, because CM DeepRx has no explicit channel estimate. Moreover, no independent clean OTA channel reference exists, so NMSE against $y/x$ would circularly treat the noisy training target as ground truth. We therefore used uncoded bit error rate (BER), measured at the first point in the receive chain where both receivers can be compared: ACM-E reaches it through ZF equalization and conventional demapping, whereas CM DeepRx outputs LLRs directly.

\begin{table}[htbp]
\caption{OFDM System and Model Parameters}
\label{tab:ofdm_params}
\centering
\begin{tabular*}{\columnwidth}{@{\extracolsep{\fill}}ll@{}}
\toprule
\textbf{System Parameter} & \textbf{Value} \\
\midrule
Carrier frequency & 5.88~GHz \\
Subcarrier spacing & 15~kHz \\
FFT size / Effective subcarriers & 256 / 192 \\
OFDM symbols per slot & 14 \\
Cyclic prefix length & 7 samples (1.82~$\mu$s) \\
Pilot symbols & Indices 2, 11; raster 8 SCs\\
Modulation (this study) & 16-QAM \\
\midrule
\textbf{Training Parameter} & \textbf{Value} \\
\midrule
Simulator & NVIDIA Sionna \cite{hoydis_sionna_2023} \\
Channel model (sim.\ training) & UMa \cite{3GPP:38.901} \\
Mobility & 0--30~m/s \\
SNR/SINR range (train / eval) & 5--40~dB / 0--40~dB \\
Optimizer & Adam ($\beta_1\!=\!0.95$) \\
\midrule
\textbf{Measurement Setup} & \textbf{Value} \\
\midrule
Environments & Urban, Suburban, Rural, Industrial \\
Mobility & Pedestrian \& Vehicular \\
Total samples & $\approx$~44\,000 \\
\bottomrule
\end{tabular*}
\end{table}

\subsection{Effect of OTA Training Data on Field Performance}

The two architectures let us compare how OTA data scaling behaves under different supervision targets and receiver structures. CM DeepRx uses bit-level supervision and is expected to benefit from increasing OTA data, whereas ACM-E uses observation-derived channel labels and may encounter a supervision-quality limit at high OTA fractions.

\begin{figure}[htbp]
\centerline{\includegraphics[width=\columnwidth]{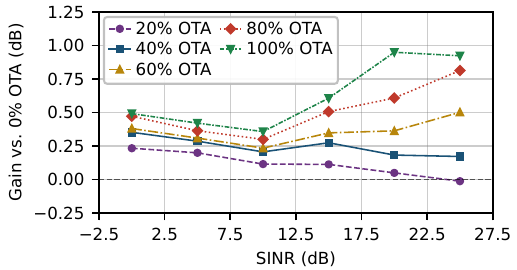}}
\caption{Equivalent SNR gain over the simulation-only baseline ($p_m = 0\%$, 100\% UMa) versus SINR for CM DeepRx at 16-QAM.}
\label{fig:snr_gain_cm}
\end{figure}

\begin{figure}[htbp]
\centerline{\includegraphics[width=\columnwidth]{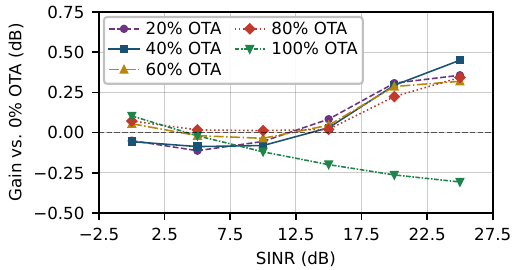}}
\caption{Equivalent SNR gain over the simulation-only baseline ($p_m = 0\%$, 100\% UMa) versus SINR for ACM-E at 16-QAM.}
\label{fig:snr_gain_acme}
\end{figure}

To evaluate this effect, both models were trained with OTA fractions $p_m \in \{0,20,40,60,80,100\}\%$. Figs.~\ref{fig:snr_gain_cm} and~\ref{fig:snr_gain_acme} show the equivalent SNR gain relative to the simulation-only baseline ($p_m=0\%$): in each SINR bin, the extra SINR the baseline needs to match the BER of the model, found by inverting the interpolated baseline BER curve.

For CM DeepRx, the gain increased monotonically with the OTA fraction in every SINR bin, and $p_m=100\%$ was best throughout, reaching about 0.9~dB at high SINR (Fig.~\ref{fig:snr_gain_cm}). With bit-level supervision, measured data add propagation and hardware diversity without introducing observation-derived channel labels (Sec.~\ref{sec:label_noise}).

ACM-E exhibited different behavior (Fig.~\ref{fig:snr_gain_acme}). Moderate OTA fractions improved performance, with $p_m=40\%$ providing the largest gain at high SINR, whereas $p_m=100\%$ degraded relative to simulation-only training. That the gap widens at high SINR, where the label-noise variance is lowest, matches the SINR dependence of the bias described in Sec.~\ref{sec:label_noise}; the highest-SINR bins also contain fewer samples and thus greater uncertainty. Overall, the behavior is consistent with OTA data reducing the simulation gap while introducing input-correlated label noise. Section~\ref{sec:label_denoising} tests the effect of changing the target construction.

\subsection{Effect of OTA Label Denoising}
\label{sec:label_denoising}

The previous results show that raw OTA supervision limits ACM-E at high OTA fractions, but do not by themselves establish causality. To probe the proposed mechanism, we constructed a denoised supervision target by applying the same separable time-frequency smoothing used in our DDCE implementation (Sec.~\ref{sec:ddce}), and repeated the mixed-training experiment. Each label $y_{ij}/x_{ij}$ was smoothed by separable uniform averaging with $(k_f,k_t)=(5,3)$. All other factors, including input data, architecture, optimization, and channel distributions, remained unchanged.

\begin{figure}[htbp]
\centerline{\includegraphics[width=\columnwidth]{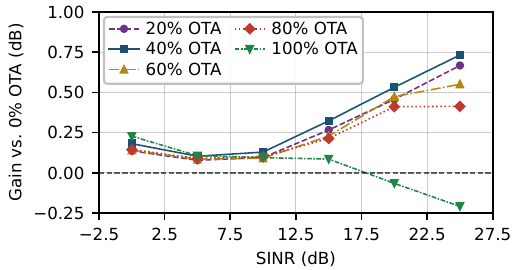}}
\caption{Equivalent SNR gain over the simulation-only baseline ($p_m = 0\%$, 100\% UMa) versus SINR for ACM-E trained with denoised OTA labels (separable uniform averaging, $k_f = 5$, $k_t = 3$) across OTA fractions $p_m$ at 16-QAM.}
\label{fig:snr_gain_denoised}
\end{figure}

Denoised labels improved ACM-E performance at every OTA fraction across the SINR range (Fig.~\ref{fig:snr_gain_denoised}). Because the OTA fractions were matched, the gain came from the supervision target rather than from additional measurement diversity. A moderate fraction ($p_m=40\%$) remained best, with fractions from 20 to $80\%$ closely matched at SINRs up to 20~dB, so we adopted the denoised $p_m=40\%$ variant as the best ACM-E configuration.

At $p_m=100\%$, denoising reduced the required SINR for BER $=10^{-2}$ from 16.07 to 15.80~dB (Table~\ref{tab:ber_sensitivity}), and the fully measured case matched or exceeded the simulation-only baseline over most of the SINR range. The remaining high-SINR gap is expected, because smoothing reduces but does not remove supervision error and introduces its own bias relative to the genie channel. Smoothing also changes label variance and imposes a channel prior, so this experiment supports the proposed mechanism but does not isolate the correlation.

To test the correlation directly, we repeated the training in simulation (UMa, two seeds) with three targets: the genie channel, the correlated label $y_{ij}/x_{ij}$, and the genie channel plus independent noise of equal power. Measured against the genie channel, independent label noise increased the NMSE by at most 0.5~dB relative to clean labels, whereas correlated labels increased it by 1.8 to 2.7~dB at 5 to 15~dB SNR. The correlated-label estimator also followed the input noise: above 20~dB SNR, its variance across input-noise realizations was 4 to 6~dB higher than with clean labels, versus at most 1.1~dB with independent label noise.

\begin{figure}[htbp]
\centerline{\includegraphics[width=\columnwidth]{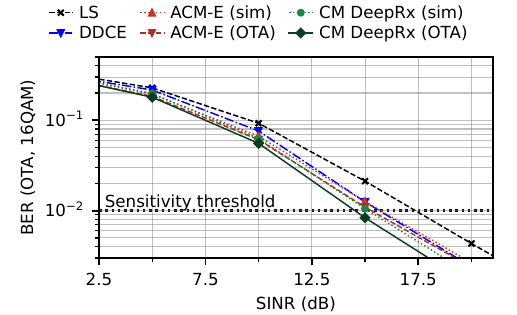}}
\caption{OTA BER versus SINR at 16-QAM for LS, DDCE, ACM-E (simulation-only and its best, denoised OTA mix at $p_m\!=\!40\%$), and CM DeepRx (simulation-only and its best OTA mix at $p_m\!=\!100\%$). The dotted line marks the sensitivity threshold at BER $=10^{-2}$.}
\label{fig:ber_sinr_combined}
\end{figure}

\subsection{Complexity and Resource Efficiency}
\label{sec:complexity}

Table~\ref{tab:complexity} summarizes computational requirements. All methods were evaluated end-to-end from $\mathbf{Y}$ to LLRs for a single OFDM slot with batch size 1; ZF equalization and demapping were included for channel-estimation methods.

Both neural models are feed-forward convolutional networks with fixed inference cost. Despite its higher FLOP count, CM DeepRx achieves latency comparable to LS because it replaces several sequential receiver stages with a single feed-forward network. ACM-E has similar latency to iterative DDCE, reflecting the cost of its encoder-decoder architecture combined with conventional receiver processing.\footnote{Latencies were averaged over 100 CUDA-synchronized runs after 10 warmup iterations on an NVIDIA GB10 using unoptimized PyTorch (FP16/FP32 for neural methods). MFLOPs were measured with \texttt{FlopCounterMode}; DDCE operations were counted manually. Values are intended for relative comparison.}

\begin{table}[htbp]
\caption{Method Comparison (single-slot, $\mathbf{Y}\to$LLR pipeline)}
\label{tab:complexity}
\centering
\begin{tabular*}{\columnwidth}{@{\extracolsep{\fill}}lccr@{}}
\toprule
\textbf{Method} & \textbf{Params} & \textbf{MFLOPs} & \textbf{ms} \\
\midrule
LS + Linear Interp.     & --    & $<1$  &  0.8 \\
DDCE ($T\!=\!3$)        & --    &  8    &  3.3 \\
ACM-E                   & 412K  & 2\,174 &  3.1 \\
CM DeepRx               & 475K  & 3\,329 &  0.9 \\
\bottomrule
\end{tabular*}
\end{table}

\subsection{Channel Estimation versus End-to-End Gains}
\label{sec:ce_vs_e2e}

At comparable parameter counts, the best CM DeepRx variant ($p_m = 100\%$) consistently outperformed the best ACM-E (denoised, $p_m = 40\%$) across the SINR range (Fig.~\ref{fig:ber_sinr_combined}), reaching BER $= 10^{-2}$ at approximately 0.8~dB lower SINR (Table~\ref{tab:ber_sensitivity}), well above the per-seed standard deviation. Several coupled differences can contribute to this gap. End-to-end optimization allows CM DeepRx to jointly compensate for residual estimation errors and hardware impairments that a modular estimate-then-demap pipeline cannot, while bit-level supervision avoids ACM-E's noisy intermediate channel labels. CM DeepRx also has a modest parameter advantage (about 15\%), a higher FLOP count, and different optimization dynamics (Table~\ref{tab:complexity}). Capacity matching reduces one confound but does not decompose these effects. The best ACM-E still clearly improved on the classical DDCE.

\begin{table}[htbp]
\caption{Required SINR at BER $= 10^{-2}$, 16-QAM. All values are in dB. SINR and gain values are means over 5 seeds; the standard deviation is given in the last column; LS and DDCE are deterministic.}
\label{tab:ber_sensitivity}
\centering
\begin{tabular*}{\columnwidth}{@{\extracolsep{\fill}}lccc@{}}
\toprule
\textbf{Method} & \textbf{SINR} & \textbf{Gain vs.\ LS} & \textbf{Std} \\
\midrule
LS (reference)                            & 17.4 & ---   & ---  \\
DDCE                                      & 16.05 & +1.35 & ---  \\
ACM-E, sim-only ($p_m\!=\!0\%$)           & 15.86 & +1.54 & 0.10 \\
ACM-E, raw OTA ($p_m\!=\!100\%$)          & 16.07 & +1.33 & 0.04 \\
ACM-E, denoised OTA ($p_m\!=\!100\%$)     & 15.80 & +1.60 & 0.08 \\
ACM-E, denoised OTA best ($p_m\!=\!40\%$) & 15.52 & +1.88 & 0.07 \\
CM DeepRx, sim-only ($p_m\!=\!0\%$)       & 15.38 & +2.02 & 0.09 \\
CM DeepRx, best OTA ($p_m\!=\!100\%$)     & 14.77 & +2.63 & 0.12 \\
\bottomrule
\end{tabular*}
\end{table}

Table~\ref{tab:ber_sensitivity} allows a partial separation of these effects at the sensitivity threshold (BER $=10^{-2}$). With simulation-only training, where both receivers have clean supervision, CM DeepRx required 0.48~dB less SINR than ACM-E, which reflects receiver structure and loss form. With fully measured training, the gap grew to 1.30~dB when ACM-E used raw OTA labels, whereas between the best configurations, with denoised labels for ACM-E, it was 0.75~dB. The supervision target therefore accounts for a substantial part of the difference.

These results point to a broader design principle that extends beyond channel estimation: OTA data are not inherently harmful, but their utility is governed by how the supervision target is extracted from the measurements. The same mechanism may arise in other tasks where supervision targets are derived from the received signal, since any target that inherits measurement noise can impose an irreducible error floor that additional data cannot remove. While this effect relates to classical errors-in-variables and correlated-label-noise settings~\cite{frenay_classification_2014}, our contribution is not to introduce that concept but to identify OTA channel supervision as an inherent instance of input-correlated supervision, derive its consequences for learned receivers, and provide supporting experimental evidence. Learned receivers should therefore favor supervision targets that remain effectively error-free under OTA collection, such as CRC-validated bits; where noisy intermediate targets are unavoidable, improving supervision quality is often more valuable than collecting additional measurements.

\subsection{Limitations} Quantitative results are tied to the specific architectures, frequencies, and environments used; consequently, the optimal $p_m$ may vary by deployment. The OTA-pool size was fixed, so the stability of this optimum under different numbers of distinct measurements was not tested; a dataset-size sweep at fixed $p_m$ remains future work. Because this study aggregates diverse conditions, the effect of specializing a receiver to a single site also remains to be explored.

Moreover, these results are limited to 16-QAM and SISO configurations. In MIMO, a single received vector under simultaneous transmission cannot identify all per-stream channel coefficients; dense labels would therefore require orthogonal pilot patterns or repeated sounding, while averaging or denoising could reduce their input-correlated noise. For higher-order QAM, divide-by-symbol labels become more heteroscedastic because low-amplitude constellation points amplify noise, while tighter decision regions can increase BER sensitivity. Whether these effects sharpen the degradation under raw OTA supervision remains to be established. Finally, as this work focuses on supervised learning, future research into self-supervised or unsupervised paradigms may alleviate the input-correlated supervision problem by avoiding explicit noisy targets.

\section{Conclusion}

We showed that the benefits of OTA training depend strongly on the supervision target. CM DeepRx benefited monotonically over the evaluated OTA fractions through effectively error-free decoded-bit supervision, whereas ACM-E reached an optimum at an intermediate OTA fraction as dense channel labels derived from measurements introduced input-correlated supervision noise. Because the received signal and the OTA channel target share the same noise realization, MSE-based channel learning retains a residual bias that additional measurements cannot remove; a controlled simulation showed that such correlated label noise is far more harmful than independent noise of equal power, and denoising the OTA labels mitigated the degradation. Future OTA learning systems should therefore use supervision targets that are reliable and independent of the network input; where noisy targets are unavoidable, improving their construction may be more valuable than collecting more measurements.

\section*{Acknowledgment}
The authors thank Mr. Geoff Tunnicliffe for facilitating the experimental transmit license for the measurement campaign.

\bibliographystyle{IEEEtran}
\bibliography{main}

\end{document}